\documentclass[cits]{PoS}
\pdfoutput=1
\usepackage{graphicx,caption,subcaption}
\usepackage{amsmath,amssymb}
\usepackage{array,blindtext}
\usepackage{verbatim}
\usepackage{amsthm}
\usepackage{amsmath}
\usepackage{enumitem}
\usepackage{graphicx}
\usepackage{amsmath,amssymb}
\usepackage{array,blindtext}
\newcommand\beq{\begin{equation}}
\newcommand\eeq{\end{equation}}
\newcommand\bea{\begin{eqnarray}}
\newcommand\eea{\end{eqnarray}}

\title{Prepotential Formulation of Lattice Gauge Theory}

\ShortTitle{Prepotential Formulation}

\author{\speaker{Indrakshi Raychowdhury} and Ramesh Anishetty\\
Institute of Mathematical Sciences, CIT Campus, Taramani, Chennai, India\\
E-mail: \email{indrakshi@imsc.res.in}, \email{ramesha@imsc.res.in}}

\abstract{ 
Within the Hamiltonian formulation of Lattice gauge theories, prepotentials, belonging to the fundamental representation of the gauge group and defined locally at each site of the lattice, enables us to construct local loop operators and loop states. We propose a set of diagrammatic rules for the action of local gauge invariant operators on arbitrary loop states. Moreover We propose a new set of fusion variables within the prepotential aproach suitable for approaching the weak coupling limit.
}

\FullConference{The 32nd International Symposium on Lattice Field Theory,\\
		23-28 June, 2014\\
		Columbia University New York, NY}
		 
\begin{document}

\section{Introduction}
Reformulating gauge theories in terms of gauge invariant loops and strings carrying fluxes  bypassing all other redundant gauge degrees of freedom is always of active interest \cite{loops,mm}. Lattice gauge theories \cite{wil,kogut} are somewhat more suited for the loop formulation because of the fact that here one directly works with the gauge-covariant link variables or holonomies (instead of the gauge field for continuum theories) as the fundamental building bocks of gauge invariant Wilson loops. 
But yet the gauge theories formulated in terms of loops suffers from too many redundant loop degrees of freedom and there exist certain constraints known as Mandelstam constraint \cite{mans} in literature solving which one would get the relevant physical degrees of freedom. Solving the Mandelstam constraint is again a truly difficult task especially at the weak coupling limit of the theory, where all possible loops of arbitrary shapes and sizes do contribute to the low energy spectrum  as the loops are nonlocal. 
A recent development in the  Hamiltonian formulation of lattice gauge theory, namely the prepotential formulation \cite{mm,pp} has shown a way to get rid of the problem of nonlocality and proliferation of loop states for any SU(N) gauge theory in arbitrary dimension. 
 
The prepotential formulation is basically a reformulation of Hamiltonian lattice gauge theory using the Schwinger Boson representation of the gauge group. This novel technique enables us to cast the gauge invariant  operators and states in a local form at each site of the lattice.
This local gauge invariant description of the theory washes away amost all  the complications associated with the loop formulation. Specifically, the Mandelstam constraints are also local in this formulation which one can solve exactly to find exact  loop basis defined locally at each site. Thus this new local description of lattice gauge theory seems to provide the best framework for any practical computation in the field of lattice gauge theory as it involves only loops which are free from any gauge redundancy and most importantly it has the  minimal description of loops defined locally at each site of the lattice. Besides strong coupling calculations the weak coupling regime becomes much more amenable and easy to handle in terms of prepotentials.

In this paper, we briefly review the prepotential formulation of SU(2) lattice gauge theory in $2+1$ dimension in section 2. Then in section 3 we discuss the local gauge invariant operators and their actions on local gauge invariant states in prepotential formulation in terms of diagrams. Finally in section 4 we introduce the fusion variables and discuss the Hamiltonian dynamics diagrammatically.
\section{Hamiltonian Formulation and the Prepotentials}
In this section we briefly review the prepotential formulation of lattice gauge theory \cite{pp}, which provides us with a platform to work with gauge invariant operators and states defined locally at each site of the lattice. 

\begin{figure}
\centering
\begin{minipage}{.7\textwidth}
  \centering
  \includegraphics[width=0.8\textwidth]{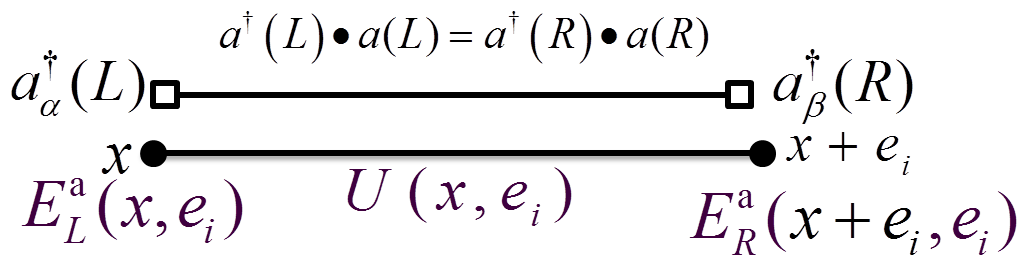}
  \captionof{figure}{Prepotentials on a link}
  \label{su2pp}
\end{minipage}%
\begin{minipage}{.3\textwidth}
  \centering
  \includegraphics[width=0.8\textwidth]{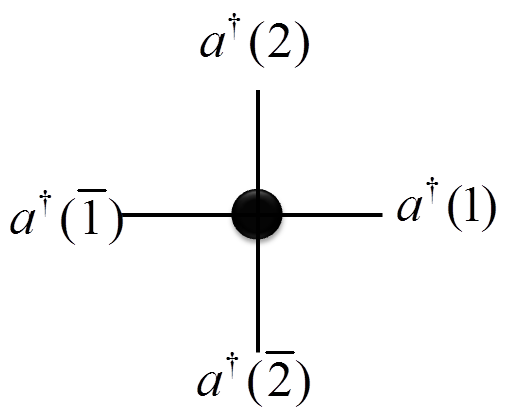}
  \captionof{figure}{A particular site on a 2 dimensional lattice and associated prepotentials}
  \label{dirnot}
\end{minipage}
\end{figure}

It is actualy a reformulation of the Kogut and Susskind \cite{kogut} Hamiltonian lattice gauge theory in which, the canonical conjugate variables are color electric fields $E^{\mathrm a}_{L/R}(x,e_i)$ defined at each site $x$, for $a=1,2,3$  with $L$ denoting the left electric field located at the starting end of the link and $R$ denotes the electric field attached at the ending point of the same link. The link operator $U(x,e_i)$'s are defined on a link originating from site  $x$ along $e_i$ direction as shown in figure \ref{su2pp}. The Hamiltonian of the theory is given by,
\bea
\hskip -1.2cm H =\underbrace{g^2\sum_{x}\sum_{\mathrm a=1}^{3}E^{\mathrm a}(x,e_i) E^{\mathrm a}(x,e_i) }_{H_{el}}
- \underbrace{\frac{1}{g^2} \sum_{\mbox{plaquette}} Tr \left(U_{\mbox{plaquette}} + U^{\dagger}_{\mbox{plaquette}}\right)}_{H_{mag}}
\label{ham}
\eea
where, $g^2$ is the coupling constant.
In (\ref{ham}),
$
U_{\mbox{plaquette}}$ is product over links around the smallest closed loop on a lattice, i.e a plaquette and 
$\mathrm a (=1,2,3)$ is the color index for SU(2).
The canonical conjugate variables satisfy the canonical commutation relation \cite{kogut}.
The gauge transformation properties of the color electric fields and link operators are as follows,
\bea
\label{eugt}
U(x,e_i)\rightarrow \Lambda(x)U(x,e_i)\Lambda^{\dagger}(x+e_i), \hskip 4cm 
\nonumber \\ 
E_L(x,e_i)\rightarrow \Lambda(x,e_i)E_L(x,e_i)\Lambda^{\dagger}(x),~~
E_R(x+e_i,e_i)\rightarrow \Lambda(x+e_i)E_R(x+e_i,e_i)\Lambda^{\dagger}(x+e_i).
\eea
Note that the left and right generators $E^\mathrm a_L(x,e_i)$ and $E^\mathrm a_R(x+e_i,e_i)$ on the link $(x,e_i)$  are not independent but are  
parallel transport of each other
$ E_R(x+e_i,e_i)= - U^{\dagger}(x,e_i) E_L(x,e_i) U(x,e_i)
$, implying the $H_{el}$ to be,
\bea
\label{E2e2}
\sum_{{\mathrm a}=1}^{3}E^{\mathrm a}(x,e_i)E^{\mathrm a}(x,e_i) \equiv 
\sum_{{\mathrm a}=1}^{3}E_L^{\mathrm a}(x,e_i)E_L^{\mathrm a}(x,e_i)
=\sum_{{\mathrm a}=1}^{3}E_R^{\mathrm a}(x+e_i,e_i)E_R^{\mathrm a}(x+e_i,e_i).
\eea
The lattice version of $SU(2)$ Gauss law 
constraint at every lattice site $n$ is
\bea
\label{gl}
\hskip 2cm G(n) = \sum_{i=1}^{d}\Big( E_L^{\mathrm a}(x,e_i) + E_R^{\mathrm a}(x+e_i,e_i)\Big)=0, \forall x. 
\eea

Now we reformulate this KS Hamiltonian formulation by replacing the canonical conjugate variables by associating a 
 set of Harmonic oscillator doublets (see figure \ref{su2pp}), or Schwinger Bosons $a_{\alpha}(x,e_i;l)$ and $a_{\alpha}^{\dagger}(x,e_i;l)$ for $\alpha =1,2$ with $l= L,R$ end of each link. We call these oscillators as prepotentials since one can construct the electric field operators as well as the link operators solely in terms of these
 with the gauge transformation properties (\ref{eugt}) as given below:
\bea
\label{sb} 
E_L^{\mathrm a}(x,e_i) &\equiv& 
a^{\dagger}(L)\frac{\sigma^{\mathrm a}}{2}a(L)~~,~~  E_R^{\mathrm a}(x+e_i,e_i) ~\equiv ~
a^{\dagger}(R)\frac{\sigma^{\mathrm a}}{2}a(R). \nonumber \\
\label{su2U}
U^{\alpha}{}_{\beta}(x,e_i)&\equiv & \frac{1}{\sqrt{\hat n+1}}\left(\tilde{a}^{\dagger\alpha}(L) \, a^{\dagger}_\beta(R) 
+ a^\alpha(L)  \, \tilde{a}_\beta(R)\right)\frac{1}{\sqrt{\hat n+1}}
\eea
Note that, we have suppressed the index $(x,e_i)$ with the prepotential operators and will do so whenever we consider one single link at a time. In (\ref{su2U}),
\bea 
\hat{n}(L) \equiv a^\dagger(L) \cdot a(L) = \hat{n}(R) \equiv a^\dagger(R) \cdot a(R) \equiv \hat{n} ~~~~~~~~~\mbox{ with, $\hat{n} \equiv \hat{n}(x,e_i)$}
\label{noc} 
\eea 
is obtained using (\ref{sb}) for (\ref{E2e2}) implying (\ref{noc}) to be an extra constraint which we call as abelian Gauss law in the literature.

Note that, this is indeed the most novel feature of prepotential formulation, where the
non-Abelian fluxes can be absorbed locally at a site and the Abelian fluxes spread along the links enabling one to construct local gauge invariant operators as well as states which we call linking operators and linking states. We discuss this issue in the next section. 

\section{Local Linking Operators and Linking States}

\begin{figure}[h]
\begin{center}
\includegraphics[width=0.5\textwidth,height=0.1\textwidth]
{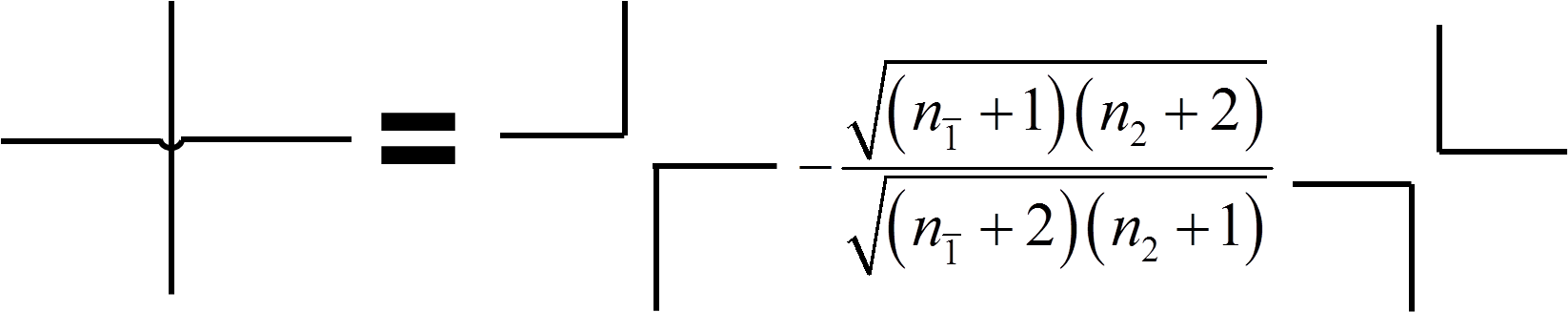}
\end{center}
\caption{Pictorial representation of Mandelstam constraint in terms of prepotentials} 
\label{mc}
\end{figure}

We now explicitly illustrate all possible local gauge invariant operators in prepotential formulation, located at each site of a 2 dimensional spatial lattice.
Concentrating at a particular site of a $2$-dimensional spatial lattice, where, $4$ links meet with each link carrying its own link operator as given in (\ref{su2U}), there exists three basic local gauge invariant operators (constructed by $U^{\alpha}{}_{\beta}(x,e_i)U^{\beta}{}_{\gamma}(x+e_i,e_j)$ at site $(x+e_i)$) which we  list below:
\bea
\hat{\mathcal O}^{i_+j_+}&\equiv & a^{\dagger}_\beta(i)\frac{1}{\sqrt{\hat n_i +1}}\frac{1}{\sqrt{\hat n_j +1}}\tilde a^{\dagger\beta}(j)\equiv \frac{1}{\sqrt{\hat n_i(\hat n_j+1)}}k_+^{ij} \label{k+} \\
\label{k+-}
\hat{\mathcal O}^{i_+j_-}&\equiv & a^\dagger_\beta(i) \frac{1}{\sqrt{\hat n_i +1}}\frac{1}{\sqrt{\hat n_j +1}} a^\beta(j) \equiv \frac{1}{\sqrt{\hat n_i}}\kappa^{ij}\frac{1}{\sqrt{(\hat n_j+2)}}\\
\label{k-}
\hat{\mathcal O}^{i_-j_-}&\equiv & \tilde a_\beta(i) \frac{1}{\sqrt{\hat n_i +1}}\frac{1}{\sqrt{\hat n_j +1}}  a^{\beta}(j) \equiv k_-^{ji}\frac{1}{\sqrt{(\hat n_i+1)(\hat n_j+2)}}~~~~~~
\eea              
where, the labels $(i/j)$ associated with prepotential operators actually denote the prepotentials associated with the links along $(i/j)$ directions at that site $x$ as per figure \ref{dirnot}. 
The maximally commuting gauge invariant set of operators $k_+^{ij}$'s are called linking operators and $k_-^{ij}$'s are their conjugates.
The linking states are constructed by the action of linking operators on strong coupling vacuum as,
\bea
|l_{ij}\rangle= \frac{\left(k_+^{ij}\right)^{l_{ij}}}{l_{ij}!}|0\rangle ~~~~\mbox{with }l_{ij}\in\mathbb Z_N
\label{lij}
\eea 
There exists six linking numbers at each site of a 2d spatial lattice characterizing a local linking state.
The number of prepotential operators (i.e eigenvalues of the operators $\hat n_i\equiv a^\dagger(i)\cdot a(i)$) at each link is counted by the linking quantum numbers in the following way:
\bea
\label{n1}
n_1 = l_{12}+l_{1\bar 1}+l_{1\bar 2}~~~,~~~
n_2 = l_{2\bar 1}+l_{2\bar 2}+l_{12}~~~,~~~
n_{\bar 1} = l_{\bar 1\bar 2}+l_{1\bar 1}+l_{2\bar 1}~~~,~~~
n_{\bar 2} = l_{1\bar 2}+l_{2\bar 2}+l_{\bar 1\bar 2}~~~~~
\eea
Note that, all the 6 linking numbers are not physical as the physical degrees of freedom for this theory is only three per lattice site implying the existence of three constraints among these linking numbers. Two of them are the abelian Gauss law constraints along two directions, i.e $n_1(x)=n_{\bar 1}(x+e_1) ~~\& ~~ n_2(x)=n_{\bar 2}(x+e_2)$. The third one is the Mandelstam constraint, which in terms of prepotentials reads as:
\bea
\label{mcor}
k_+^{1\bar 1}k_+^{2\bar 2}=k_+^{1\bar 2}k_+^{2\bar 1}- k_+^{12}k_+^{\bar 1\bar 2}
\eea
Pictorially the Mandelstam constraint is illustrated in figure \ref{mc} which allows us to choose the physical loop basis consisting  of non-intersecting loops.

Now, the action of the linking operators defined in (\ref{k+}), (\ref{k+-}) and (\ref{k-}) on the linking states defined in (\ref{lij}) are obtained as \cite{newpaper},
\bea
\label{ac++}
\hat{\mathcal O}^{i_+j_+}|\{l\}\rangle &\equiv &  \frac{(l_{ij}+1)}{\sqrt{( n_i+1)( n_j+2)}}|l_{ij}+1\rangle\nonumber \\
\hat{\mathcal O}^{i_+j_-}|\{l\}\rangle &\equiv &  \frac{1}{\sqrt{(n_i+1)(n_j+2)}}\sum_{k\ne i,j} (-1)^{S_{ik}}(l_{ik}+1) |l_{jk}-1,l_{{ik}}+1\rangle \\
\hat{\mathcal O}^{i_-j_-}|\{l\}\rangle &=& \frac{1}{\sqrt{(n_i+1)(n_j+2)}}\Bigg[ (n_i+n_j-l_{ij}+1)|l_{ij}-1\rangle +  \sum_{i', j'\{\ne i,j\}}(l_{i'j'}+1)(-1)^{S_{i'j'}}| l_{ii'}-1,l_{jj'}-1,l_{i' j'}+1\rangle  \Bigg]~~~~~~\nonumber
\eea
with $
S_{ik}=1~~\mbox{if }i>k~~\&~~S_{ik}=0~~\mbox{if }i<k.$ In \cite{newpaper} we have prescribed a set of diagrametic rules, following which the local action of gauge invariant operators on linking states as given in (\ref{ac++}) are represented pictorially in figures 4 (a), (b) and (c) respectively.  This set of rules given in \cite{newpaper} allows us to read-off the complete mathematical expression given in the above set of equations from the respective figures. This set of diagrammatic rules are extremely useful for computation of Hamiltonian dynamics in both the weak or strong coupling regime and is explicitly given in \cite{newpaper}. 

\begin{figure}
\label{acfig}
\centering
\begin{subfigure}{.17\textwidth}
  \centering
  \includegraphics[width=0.8\textwidth]{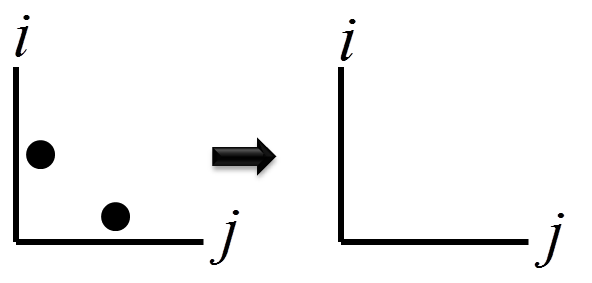}
  \caption{}{}
\end{subfigure}%
\begin{subfigure}{.33\textwidth}
  \centering
  \includegraphics[width=0.8\textwidth]{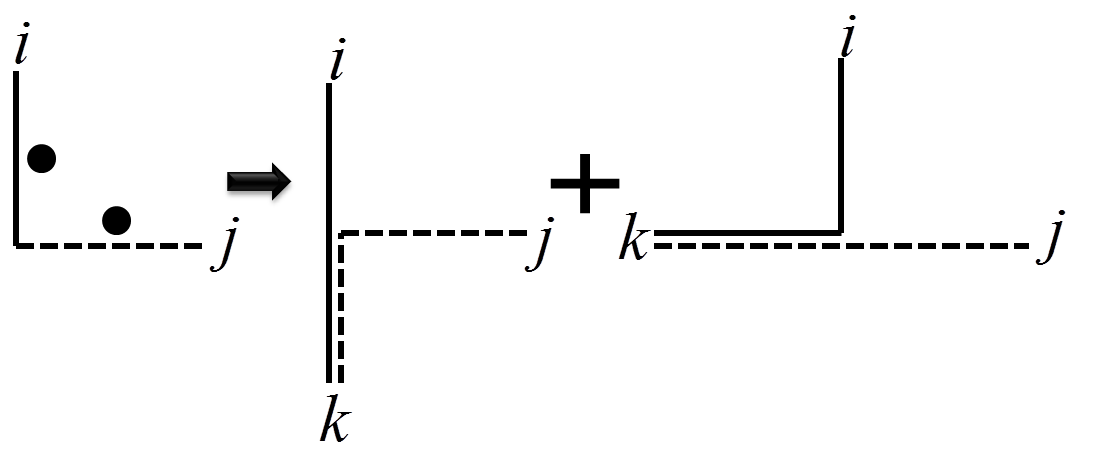}
  \caption{}{}
\end{subfigure}
\begin{subfigure}{.49\textwidth}
  \centering
  \includegraphics[width=0.8\textwidth]{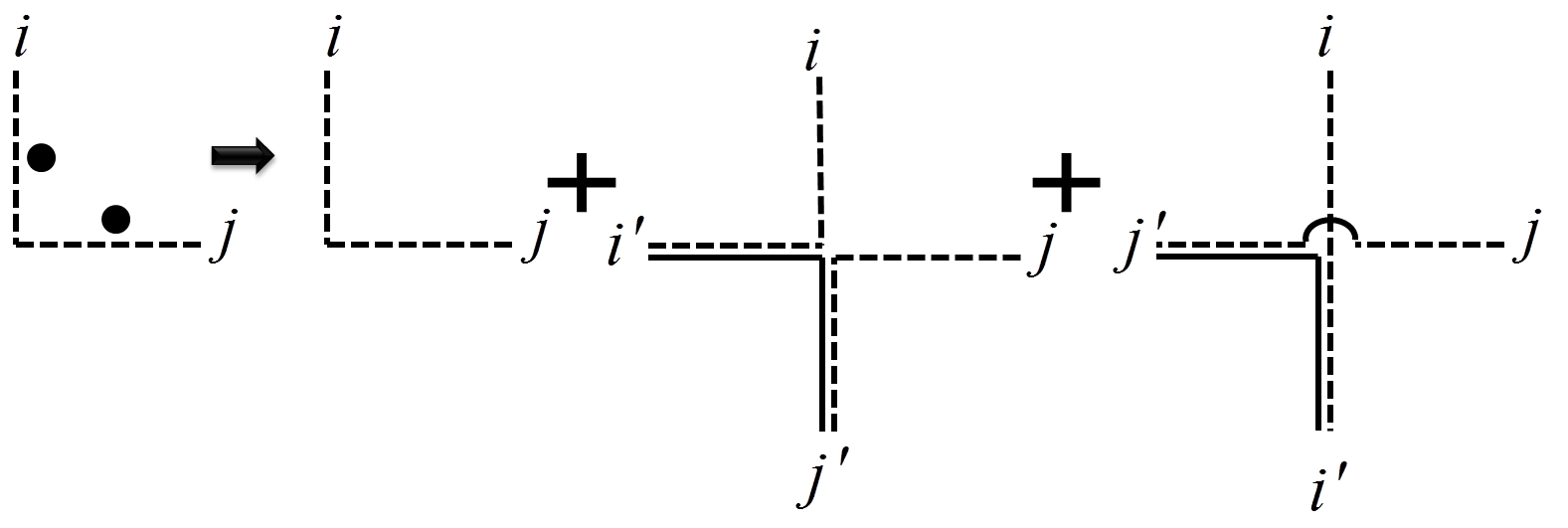}
  \caption{}{}
\end{subfigure}
\caption{Linking state produced by the action of local gauge invariant operators with number of prepotentials (a) increasing, (c) decreasing along both directions and (b) increasing along one direction and decreasing along other. Note that, the third diagram in the R.H.S of (c) vanishes for any loop state which satisfies the Mandelstam constraints.}
\end{figure}
\section{Fusion variables and Hamiltonian Dynamics}
\begin{figure}
\centering
\begin{minipage}{.5\textwidth}
  \centering
  \includegraphics[width=0.8\textwidth]{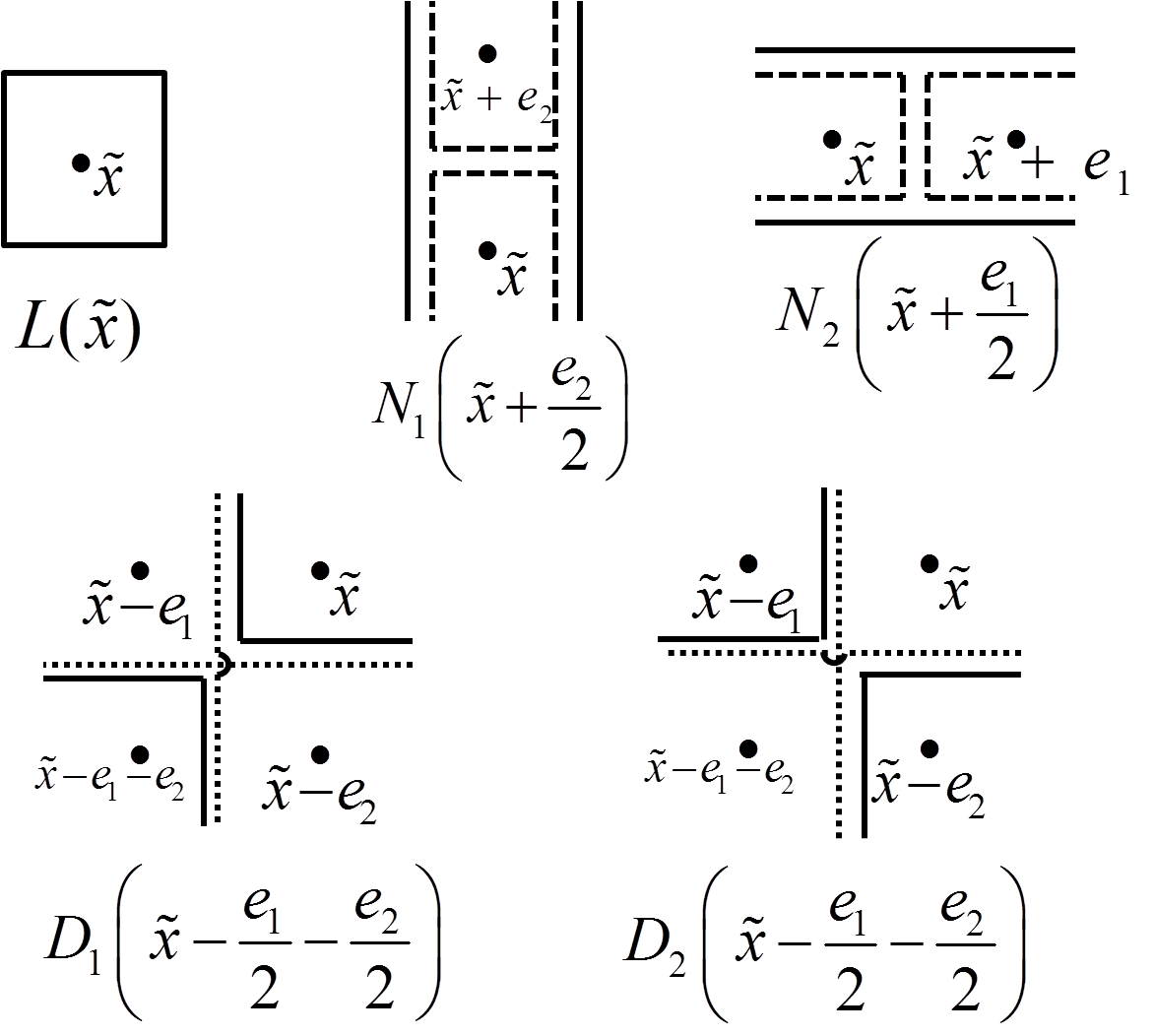}
  \captionof{figure}{Fusion variables defined at or around the dual site}
  \label{fc}
\end{minipage}%
\begin{minipage}{.5\textwidth}
  \centering
  \includegraphics[width=0.9\textwidth]{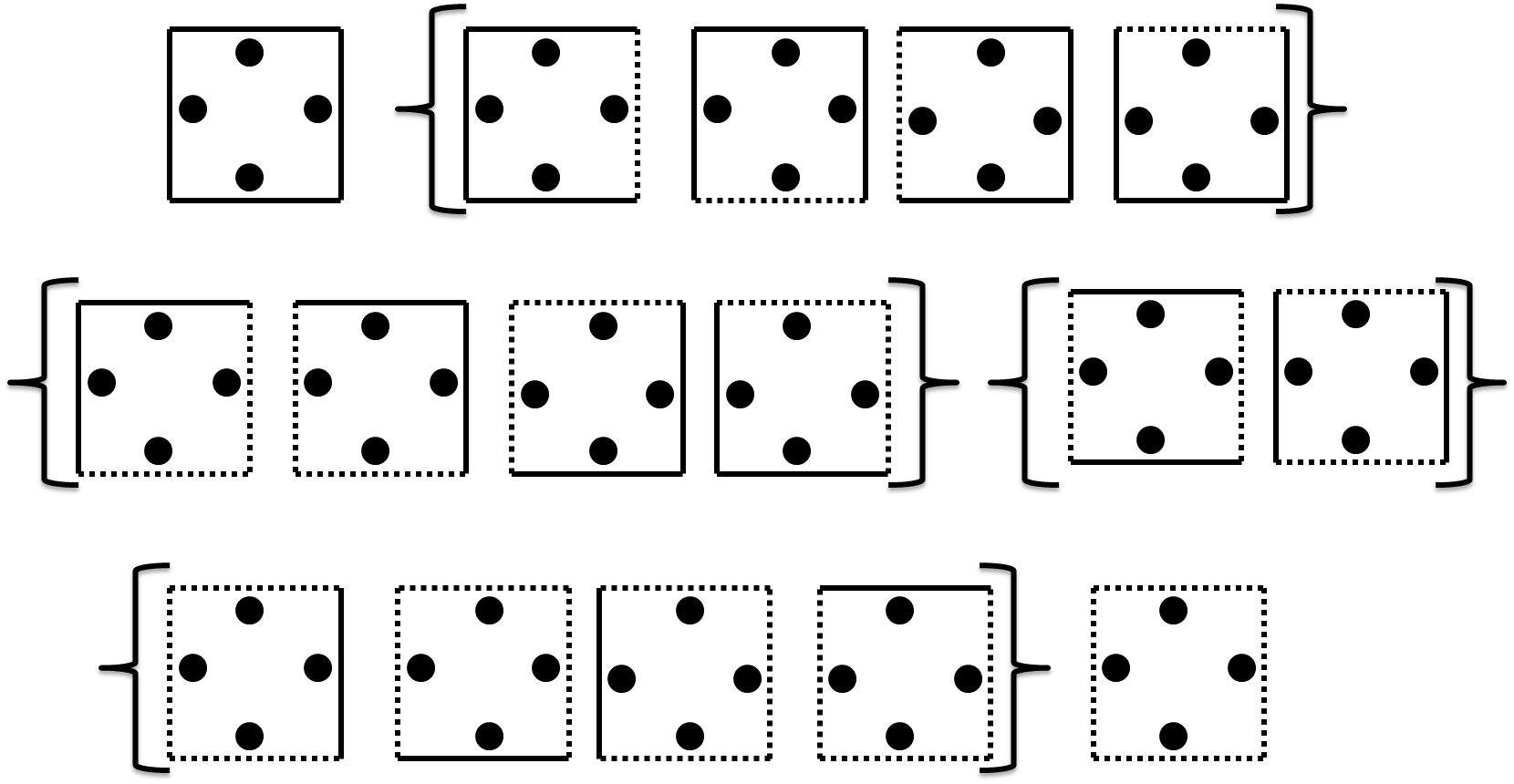}
  \captionof{figure}{The Hamiltonian operator}
  \label{h16}
\end{minipage}
\end{figure}

We introduce \cite{newpaper} a new set of fusion variables $\{L,N_1,N_2,D_1,D_2\}$ in order to characterize the loop basis as shown in figure \ref{fc}. Any loop state, characterized by  linking numbers locally at each site can as well be characterized by the set of fusion quantum numbers described pictorially in figure \ref{fc}. The fusion quantum numbers has one to one correspondence to the local linking numbers in characterization of a loop state. In fact each linking number at any site can be easily read off from the fusion quantum numbers described in figure \ref{fc} as below 
\bea
\label{mr1}
l_{12}(x)&=& L(\tilde x)-N_2(\tilde x-\frac{e_1}{2})-N_1(\tilde x-\frac{e_2}{2})+D_1(\tilde x-\frac{e_1}{2}-\frac{e_2}{2})\ge 0\\
\label{mr2}
l_{1\bar 1}(x)&=& N_2(\tilde x-\frac{e_1}{2})+ N_2(\tilde x-\frac{e_1}{2}-e_2)-D_1(\tilde x-\frac{e_1}{2}-\frac{e_2}{2})-D_2(\tilde x-\frac{e_1}{2}-\frac{e_2}{2})\ge 0\\
\label{mr3}
l_{1\bar 2}(x)&=& L(\tilde x-e_2)-N_2(\tilde x-\frac{e_1}{2}-e_2)-N_1(\tilde x-\frac{e_2}{2})+D_2(\tilde x-\frac{e_1}{2}-\frac{e_2}{2})\ge 0\\
\label{mr4}
l_{2\bar 1}(x)&=& L(\tilde x-e_1)-N_2(\tilde x-\frac{e_1}{2})-N_1(\tilde x-e_1-\frac{e_2}{2})+D_2(\tilde x-\frac{e_1}{2}-\frac{e_2}{2})\ge 0\\
\label{mr5}
l_{2\bar 2}(x)&=&N_1(\tilde x-\frac{e_2}{2})+ N_1(\tilde x-e_1-\frac{e_2}{2})-D_1(\tilde x-\frac{e_1}{2}-\frac{e_2}{2})-D_2(\tilde x-\frac{e_1}{2}-\frac{e_2}{2})\ge 0\\
\label{mr6}
l_{\bar 1\bar 2}(x)&=& L(\tilde x-e_1-e_2)-N_2(\tilde x-\frac{e_1}{2}-e_2)-N_1(\tilde x-e_1-\frac{e_2}{2})+D_1(\tilde x-\frac{e_1}{2}--\frac{e_2}{2})\ge 0
\eea
where $\tilde x$ is a dual lattice site, and $e_1$ and $e_2$ are two unit vectors along $1~\&~2$ directions of a lattice. Note that, the fusion variables $\in \mathbb Z$, provided the linking numbers are positive semidefinite as given in (\ref{mr1})-(\ref{mr6}). 

Now, concentrating on the Hamiltonian, specifically the magnetic part of the Hamiltonian (i.e the $H_{mag}$ in (\ref{ham})) and using the prepotential construction of the link variables given in (\ref{su2U}) we find \cite{newpaper} the magnetic Hamiltonian operator to be a sum of sixteen plaquette operators as shown in figure \ref{h16} diagrammatically. In figure \ref{h16}, the solid line along a link denotes the presence of prepotential creation operator on that link whereas, a dotted line denotes the annihilation operators on that. Clearly the whole set is rotationally symmetric and hermitian. Note that, each of these 16 plaquette operators is a product of $4$ local gauge invariant operators at its four vertices. These local gauge invariant operators are of all possible type of local gauge invariant operators defined in (\ref{k+})-(\ref{k-}). Using the local loop actions illustrated in figure 4 one can find the complete dynamics of the loops under this magnetic  Hamiltonian. Moreover, the fusion variables are extremely suited to describe this dynamics and the Hamiltonian operator can be completely rewritten in terms of the shift operators (defined in \cite{newpaper}) in the fusion variables. 
 
\section{Future Directions}
The prepotential formulation enables us to describe the theory completely in terms of gauge invariant variables locally defined at each site. Further the diagrammatic rules prescribed enables one to compute the complete loop dynamics diagrammaticaly without going into the detail of algebraic computation and dealing with Clebsch Gordon coefficients. Moreover the fusion variables enables us to enumerate all possible loop state states by specifying a set of integers throughout the lattice which seems to be extremely useful in dealing the loop formulation numerically. The most important feature of the fusion variables is that, among the five variables defined in figure \ref{fc}, the first one, i.e $L(\tilde x)$ is the basic loop variable which acts as the primary building blocks for any arbitrary loop. In fact the loop states are primarily consisting of plaquette loops or $L(\tilde x)$, and then the neighbouring plaquettes are connected by the other four fusion variables in all possible way in order to get all possible loops in the theory. Hence, in the strong coupling limit of the theory where only smaller loops with small fluxes contribute to finite energy we have $L(\tilde x)\rightarrow 0$  and for weak coupling limit $L(\tilde x)\rightarrow \infty$, $\forall x$. Thus these new set of variables becomes extremely useful to approach both the limit analytically as well as numerically. 

Moreover, the recently developed tensor network approach to Hamiltonian lattice gauge theory \cite{tensor1} should find this loop formulation most suitable to proceed  with for non Abelian gauge theories. This loop formulation and diagrammatic techniques should also be extremely useful towards the  aim of the construction of quantum simulations \cite{qs} for lattice gauge theories.

\end{document}